\documentstyle[preprint,aps]{revtex}
\newcommand{\bm}[1]{\bbox{#1}}
\renewcommand{\theequation}{\arabic{section}.\arabic{equation}}

\tighten
\begin{document}


\draft

\title{\bf An extended massless phase and the Haldane phase in a spin-1
isotropic antiferromagnetic chain}
\author{Chigak Itoi and   Masa-Hide Kato}
\address{Department of Physics, College of Science and Technology, Nihon 
 University. \\ Kanda Surugadai, Chiyoda-ku, Tokyo 101, Japan.}
\date{\today}

\maketitle

\begin{abstract}
We study the phase diagram of isotropic spin-1 models
in the vicinity of the Uimin-Lai-Sutherland (ULS) model.
This is done with the help of  a level-one $\text SU(3)$
Wess-Zumino-Witten model with certain marginal
perturbations.
We find that the renormalization group flow has infrared stable and unstable 

trajectories divided by a critical line
on which the ULS model is located.
The infrared unstable trajectory produced by a marginally relevant 
perturbation
generates an exponential mass gap for the Haldane phase,
and thus the universality class of the
transition from the massless phase to
the Haldane phase at ULS point
is identified with the Berezinski\u\i-Kosterlitz-Thouless type.
Our results support recent numerical studies by  F\'ath and S\'olyom.
In  the massless phase, we calculate logarithmic finite-size
corrections of the energy for the $\text SU(\nu)$-symmetric
and asymmetric models.
\\
\end{abstract}


\section{Introduction}
The phase diagram of isotropic spin-1 chains
has not yet been understood sufficiently.
The characteristics of  ground states can change
drastically depending on a coupling constant of the model \cite{SJG}.
Even though there are many rigorous \cite{AKLT,KeTa} and exact
\cite{ULS,KR,TB,BB} works
at several isolated points, one encounters
unconformable issues in a certain region,
especially in the non-integrable region
around an integrable point of the Uimin-Lai-Sutherland (ULS) model.

The general form of the spin-1 Hamiltonian
which consists of nearest neighbor interactions
with rotational symmetry is
\begin{equation}
H(\theta) =  \sum_{j=1}^{L} \left[ \cos{\theta}
\left( \bm{S}^{\ }_{j} \cdot \bm{S}^{\ }_{j+1}\right)
+ \sin{\theta}\left( \bm{S}^{\ }_{j} \cdot
\bm{S}^{\ }_{j+1} \right)^2 \right],
\label{Hamil}
\end{equation}
where the coupling constant is controlled
by one parameter ${ \theta}\in[0,2\pi)$.
It is our main concern to understand the macroscopic behavior
in the vicinity of the ULS point $\theta = \pi/4$
\cite{ULS,KR}.
It is known that
the ULS model has massless excitations described by
the Wess-Zumino-Witten (WZW) model.
The region $|\theta| < \pi/4$, which contains the standard Heiseberg 
antiferrromagnet ($\theta=0$), is believed to be
in the Haldane phase which has only massive excitations,
as suggested by some numerical works \cite{NB,Nomu} and rigorous studies at 

$\theta=\tan^{\!-1}{(1/3)}$ \cite{AKLT,KeTa}.
On the other hand, the natures of the model in the region $\pi/2 > \theta > 
\pi/4$  are  theoretically less understood.

In this paper employing a renormalization group (RG) method in a
continuum field theory,
we show that the region $\theta > \pi/4$ near
$\theta = \pi/4$ is  a massless phase, and that
the phase transition from this massless phase to the Haldane phase at
ULS point belongs to the Berezinski\u\i-Kosterlitz-Thouless (BKT) type
universality class. This result is consistent with a numerical study 
obtained
by F\'ath and S\'olyom \cite{FS}.
For this purpose, we map  the
ULS model to the $\text SU(3)_1$ WZW model, which
reproduces some exact results obtained
from the Bethe ansatz \cite{ULS,KR,Jo,PT,LS,Sch,MNTT}.
In the non-integrable region around the ULS point,
we show that the $\text SU(3)_1$ WZW model is perturbed
by adding a $\text SU(3)$-breaking marginal operator
which causes the BKT transition.
We observe several nontrivial behaviors as
in some other CFT deformed by marginal operators \cite{AGSZ,EAT,HHM,Eg}.
Despite a number of studies on
the BKT transition
and the logarithmic corrections in ${\text SU}(2)$ systems,
those concerned with ${\text SU}(\nu)$ symmetry for
$\nu > 2$ has been seldom discussed.
Here, we study the BKT transition and
the logarithmic correction in
the $\nu > 2$ case and we find its different nature
from $\nu=2$ case. The obtained
continuum theory enables us to calculate the logarithmic finite-size
correction in the energy of the ground state and the first excited states
in the region of the massless phase. Following Ludwig and Cardy 
\cite{LC,Ca},
the finite size correction to the ground state energy of the model
in a strip space with the width $L$ is
\begin{equation}
\begin{array}{@{\,}c@{\ }c@{\ }l}
{\cal E}^{\ }_{\rm G.S} &=& \varepsilon_{\infty}^{\ }
L \displaystyle  -{\pi v \over 6L} c(L),
\medskip \\ c(L) &=& \displaystyle c_{\rm vir}
+ \frac{d_{\rm G.S}}{(\ln L)^3} +
O\left( \frac{\ln{(\ln{L})}}{(\ln{L})^4},\frac{1}{(\ln{L})^4} \right)
\ ,
\label{Genergy}
\end{array}
\end{equation}
where $\varepsilon_{\infty}^{\ }$ is the non-universal
bulk contribution to the ground state energy
depending on cut-off scale.
The minimal energy of an excited state related
to a certain primary field with conformal weight $x_{n}^{\ }/2$ is given by 

\begin{equation}
\begin{array}{@{\,}c@{\ }c@{\ }l}
{\cal E}^{\ }_{n} &=& {\cal E}^{\ }_{\rm G.S}
+ \displaystyle {2 \pi v \over L} \gamma_n^{\ }(L),\medskip \\
 \gamma_n^{\ }(L) &=&  \displaystyle x_n^{\ }
 + {d^{\ }_{n} \over \ \ln L\ }
 + O\left( \frac{\ln{(\ln{L})}}{(\ln{L})^2}, \frac{1}{(\ln L)^2}\right)\ , 

\label{Exenergy}
\end{array}
\end{equation}
where  $d_{n}^{\ }$ is a coefficient of
a certain three point function. We calculate these universal coefficients
of the logarithmic corrections by the obtained continuum field theory.

The outline of this paper is as follows.
In Sec. \ref{sec:critical}, a strong coupling abelian gauge theory
is introduced as a critical field theory of the ULS model,
which allows us to evaluate exact values of universal quantities.
We show the equivalence of this critical theory to
the level-one ${\text SU}(3)$ WZW model.
This argument can be
generalized to a certain ${\text SU}(\nu)$
symmetric spin model which includes the ULS model in the $\nu = 3$ case.
In Sec. \ref{sec:marginal},
we discuss an extended non-integrable spin model with
$\text SU(\nu)$-asymmetric interaction
on the basis of the level-one $\text SU(\nu)$ WZW model with
an asymmetric perturbation.
We pin down the marginal operator
$\sum_{A=1}^{\nu^2-1}{\cal J}^{A}_{\!{\scriptscriptstyle L}{\alpha 
\beta}}(z)
{\cal J}^{A}_{\!{\scriptscriptstyle R}{\alpha \beta}}(\overline z)$
in the $\text SU(\nu)_{1}$ WZW model
as the $\text SU(\nu)$-asymmetric
interaction in the original spin model.

The logarithmic corrections of its energy
in the massless phase are evaluated and the difference between
$\text SU(3)$-symmetric and asymmetric model is indicated.
Finaly, we discuss the universality class of the transition
from the massless phase to the Haldane phase which
belongs to the BKT type.

\setcounter{equation}{0}
\section{Critical theory of SU($\nu$) spin chain}
\label{sec:critical}
To  begin with, we extend
a $\rm QED_{2}$ description for
$\text SU(2)$ spin model \cite{IM} to the  $\text SU(\nu)$ one.
The spin chain is mapped to the WZW model with some perturbations
by this method.

We redefine the Hamiltonian (\ref{Hamil}) near $\theta=\pi/4$ as
\begin{equation}
H(\gamma) \equiv {1\over {\cos{\theta}} } H(\theta)  =\sum_{j=1}^{L} \left[ 
\left( \bm{S}^{\ }_{j} \cdot \bm{S}^{\ }_{j+1}\right) + \gamma \left( 
\bm{S}^{\ }_{j} \cdot \bm{S}^{\ }_{j+1} \right)^2 \right]
\label{redefH}
\end{equation}
with $\gamma=\tan{\theta}$.
We use fermion operators $c_{j \alpha}, c_{j \alpha} ^\dagger$
for the spin variables
\begin{equation}
\bm{S}_{j} = \sum_{\alpha,\beta = 1}^{3} c^{\dagger}_{j,\alpha}(\bm{L})^{\ 
}_{\alpha\beta}c^{\ }_{j,\beta},
\label{SPINop}
\end{equation}
where $ {L}^{x},{L}^{y}$ and ${L}^{z}$ are spin-1 matrices.
In this case, eq(\ref{redefH}) can be expressed in the fermions \cite{Af}
\begin{equation}
H(\gamma) =\sum_{j=1}^{L} \left[ c^{\dagger}_{j,\alpha}c^{\ 
}_{j,\beta}c^{\dagger}_{j+1,\beta}c^{\ 
}_{j+1,\alpha}+(\gamma-1)c^{\dagger}_{j,\alpha}c^{\ 
}_{j,\beta}c^{\dagger}_{j+1,\alpha}c^{\ }_{j+1,\beta} \right],
\label{FHamil}
\end{equation}
in which a trivial constant is neglected.
Here the local constraint, $\sum_{\alpha=1}^{3}$ 
  $c^{\dagger}_{j,\alpha}c^{\ }_{j,\alpha}$ $=1$, is imposed in order to 
restrict the dimension of  physical space to three at each lattice site.
Due to this constraint, empty, double and triple occupancy
states of the fermions are forbidden at each lattice site.
The first term in eq(\ref{FHamil}) is a exchange operator
between nearest neibour sites and
the second one is a projector onto a singlet bond.
Eq(\ref{FHamil}) has  the local U(1) symmetry,
 a translational symmetry by one lattice site
for all values of $\gamma$. In addition,
a global $\text SU(3)$ symmetry appears at the ULS point ($\gamma=1$)
where
the Hamiltonian (\ref{FHamil})consists of only
exchange operators, and the model becomes Bethe ansatz solvable.
This fermion expression can be extended
to Bethe ansatz solvable model with higher spin.
When spin-$S$,
$2S+1$ kinds of fermion on each lattice site are introduced,
and the constraint on each site is given
by $\sum_{\alpha=1}^{2S+1}c^{\dagger}_{j,\alpha}c^{\ }_{j,\alpha}=1$.
In general, bond interactions of an isotropic spin-$S$ chain are represented 

by a polynomial of ${\text X}= \bm{S}_i \cdot \bm{S}_j$.
These integral families for higher spin chains are classified by
Batchelor, Yung and Kennedy \cite{BY}. One of those is the ULS model with 
an
arbitrary spin. We give
the expressions of the exchange operator in terms of spin matrix
for $S \leq 2 $ in Table \ref{Table1}.
Hereafter we discuss the fermionized Hamiltonian (\ref{FHamil})
as $\nu$ species ($\nu=2S+1$).

The  euclidean action is written by
introducing Lagrangian multiplier $\chi$ for the local constraint
and the Hubbard-Stratonovich
transformation
\begin{equation}
{\cal A}^{\ }_{\gamma} =\int^{\beta}_{0} \! \!d\tau
\sum_{j=1}^{L}
\left( c^{\dagger}_{j,\alpha}{\partial^{\ }_\tau} c^{\ }_{j,\alpha} +
{\text i} \chi^{\ }_{j} ( c^{\dagger}_{j,\alpha} c^{\ }_{j,\alpha}-1 )
+{\cal H}_{j,j+1}(\gamma) \right) ,
\label{Laction}
\end{equation}
where $\beta$ is an inverse temperature.
The Hamiltonian for one bond interaction  is
$$
{\cal H}_{j,j+1}(\gamma)={\cal Q}^{\ast}_{j,j+1}{\cal Q}^{\ 
}_{j,j+1}-c^{\dagger}_{j,\alpha}{\cal Q}^{\ }_{j,j+1}c^{\ }_{j+1,\alpha}
-c^{\dagger}_{j+1,\alpha}{\cal Q}^{\ast}_{j,j+1}c^{\ }_{j,\alpha}
 +(\gamma-1)c^{\dagger}_{j,\alpha}c^{\ 
}_{j,\beta}c^{\dagger}_{j+1,\alpha}c^{\ }_{j+1,\beta}.
\label{QHamil}
$$
The complex auxiliary fields
$\{{\cal Q}^{\ }_{j,j'}\},\{{\cal Q}^{\dagger}_{j,j'}\}$
are introduced to decompose the two-body fermion interaction into
single body.
A local U(1) gauge transformation
\begin{equation}
c_{j,\alpha} \rightarrow  e_{\ }^{{\text i}\varphi_{j}}c_{j,\alpha},
\quad
\chi_j \rightarrow \chi_j -\partial_{\tau} \varphi_j,
\quad
{\cal Q}_{j,j'} \rightarrow e^{{\text i} \varphi_{j}}
{\cal Q}_{j,j'} e^{-{\text i} \varphi_{j+1}},
\label{gaugetransform}
\end{equation}
preserves eq(\ref{Laction}).

First, we study $\text SU(\nu)$-symmetric point $\gamma = 1$ by the mean 
field theory
without taking into account the local constraint
$\sum_{\alpha=1}^{\nu}c^{\dagger}_{j,\alpha}c^{\ }_{j,\alpha}=1$.
We shall treat the
local constraint later.
In the  mean field theory, the auxiliary field $Q_{j,j'}$
is a constant $R_{0}$, 
and then dispersion relation becomes
$\varepsilon(k)=- R^{\ }_0 \cos{ka}$, where $a$ is a lattice spacing.
The ground state is given by the fermi sea filled up
to fermi level $\pm k_{F}$ with $k_{F}=\pi/\nu a$.
The low energy physics can be described in terms of
$\psi^{\ }_{\!\scriptscriptstyle L}$
and $\psi^{\ }_{\!\scriptscriptstyle R}$
which is the lattice fermion operator $c_{j,\alpha}$
only around the fermi surface with a certain low energy
cutoff $ \Lambda (<< k_F)$
$\pm k_{F}$ as
\begin{equation}
{1\over\sqrt{a}} c_{j,\alpha} \simeq \psi^{\ }_{\!{\scriptscriptstyle 
 L}\alpha}(x) \exp{(-{\text i}k_{F}x)}+{\psi}^{\ }_{\!{\scriptscriptstyle 
 R}\alpha}(x)\exp{({\text i}k_{F}x)},\quad x\equiv j\,a .
\label{Lfermion}
\end{equation}
In this representation, the local gauge transformation
eq(\ref{gaugetransform})
corresponds to the U(1) vector transformation
$$
\psi^{\ }_{\!{\scriptscriptstyle  L}\alpha}(x) \rightarrow
e^{\varphi(x)} \psi^{\ }_{\!{\scriptscriptstyle  L}\alpha}(x),
$$
$$
\psi^{\ }_{\!{\scriptscriptstyle  R}\alpha}(x) \rightarrow
e^{\varphi(x)} \psi^{\ }_{\!{\scriptscriptstyle  R}\alpha}(x).
$$
A translation by one-site on the original lattice space,
$$
c_{j,\alpha}\rightarrow c_{j+1,\alpha}\exp({\text i}k_{F}a),
$$
corresponds to a chiral $\bm{Z}_\nu$ transformation:
$$
\left( \psi^{\ }_{\!{\scriptscriptstyle  L}\alpha}, \psi^{\ 
}_{\!{\scriptscriptstyle  R}\alpha} \right)\rightarrow \left(\psi^{\ 
}_{\!{\scriptscriptstyle  L}\alpha},\psi^{\ }_{\!{\scriptscriptstyle 
 R}\alpha}
\exp{(2{\text i}k_{F}a)} \right).
$$
As far as the translational symmetry is not broken,
the effective field theory becomes chiral $\bm{Z}_\nu$-invariant.

Now, we take into account the deviation from the
mean field approximation.
In the following parametrization of the auxiliary field
$$
\textstyle
{\cal Q}_{j,j'}= R({j+j'\over 2})\exp{\left\{ {\text i} |j-j'| 
A_{1}({|j+j'|\over 2}) \right\} },
$$
the deviation of ${\cal Q}_{j,j'}$ becomes
\begin{equation}
\textstyle
{\cal Q}_{j,j'}\simeq R^{\ }_0 + \delta R({j+j'\over 2}) +{\text i} |j-j'| 
A_{1}({|j+j'|\over 2}).
\label{approQ}
\end{equation}
The local constraint is expressed as{
\setcounter{enumi}{\value{equation}}
\addtocounter{enumi}{1}
\setcounter{equation}{0}
\renewcommand{\theequation}{\arabic{section}.\theenumi\alph{equation}}
\begin{equation}
\psi_{\!{\scriptscriptstyle L} \alpha}^\dagger
\psi_{\!{\scriptscriptstyle L} \alpha}^{\ }(x) = 0,
\quad  \psi_{\!{\scriptscriptstyle R} \alpha}^\dagger
\psi_{\!{\scriptscriptstyle R} \alpha}^{\ }(x) =0,
\label{constraint1}
\end{equation}
\begin{equation}
\psi_{\!{\scriptscriptstyle L} \alpha}^\dagger
\psi_{\!{\scriptscriptstyle R} \alpha}^{\ }(x) +
\psi_{\!{\scriptscriptstyle R} \alpha}^\dagger
\psi_{\!{\scriptscriptstyle L} \alpha}^{\ }(x) = 0.
\label{constraint2}
\end{equation}
\setcounter{equation}{\value{enumi}}
}
We obtain a chiral $\bm{Z}_\nu$-invariant effective
Lagrangian in terms of low energy variables:
$$
{\cal L} = {\cal L}^{\ }_{0} +{\cal L}^{\ }_{{\rm int}},
$$
where
\begin{equation}
\begin{array}{@{\,}c@{\ }c@{\ }l}
{\cal L}^{\ }_{0} &=& 2\psi^{\dagger}_{\!{\scriptscriptstyle L}\alpha} 
\left( \overline{\partial}+{\text i}\overline{A} \right)\psi^{\ 
}_{\!{\scriptscriptstyle  L}\alpha}
+2\psi^{\dagger}_{\!{\scriptscriptstyle  R}\alpha}  \left( \partial+{\text 
i} A \right)\psi^{\ }_{\!{\scriptscriptstyle  R}\alpha}, \medskip \\
{\cal L}^{\ }_{\rm int} &=& {\rm constant} \times 
\psi^{\dagger}_{\!{\scriptscriptstyle  L}\alpha}\psi^{\ 
}_{\!{\scriptscriptstyle  L}\beta}\psi^{\dagger}_{\!{\scriptscriptstyle 
 R}\beta}\psi^{\ }_{\!{\scriptscriptstyle  R}\alpha}.
\end{array}
\label{Caction}
\end{equation}
Here, the gauge field $A_0$ is
a low energy variable corresponding to the Lagrangian multiplier
$\chi = a A_0 $ 
and  $A = A_0 + {\text i}A_1, \overline{A} =A_0 - {\text i} A_1$.
This effective theory is a perturbed abelian gauge field theory
with a sound velocity $v = R_0 a \sin{(k_{F}a)}$, here $v$ is set to 
unity.
In deriving the Lagrangian ${\cal L}$ we have picked up the terms to
$O(a^2)$ and neglected the highly oscillating
terms and higher derivative terms.
The four-fermi interaction ${\cal L}^{\ }_{\rm int}$ is
induced by performing the gaussian integration over
the $\delta R_0$-field and also by
the second constraint (\ref{constraint2}).
This interaction can be expressed in the form
$$
{1 \over \nu} j^{\ }_{\!{\scriptscriptstyle L}}(z){j}^{\ 
}_{\!{\scriptscriptstyle R}}(\overline{z}) + 2\sum_{A=1}^{\nu^2-1}{\cal 
J}_{\!{\scriptscriptstyle L}}^A (z){\cal J}_{\!{\scriptscriptstyle 
R}}^A(\overline{z}),
$$
where
$j^{\ }_{\!{\scriptscriptstyle L(R)}}=$
$\psi^{\dagger}_{\!{\scriptscriptstyle  L (R)}\alpha}
\psi^{\ }_{\!{\scriptscriptstyle  L (R)}\alpha}$  and
${\cal J}_{\! \scriptscriptstyle L (R) }^A =
\psi^{\dagger}_{\!{\scriptscriptstyle  L (R)}\alpha}
{T}^{A}_{\alpha\, \beta} \psi^{\ }_{\!{\scriptscriptstyle  L  (R)}\beta}$. 

In Appendix A, the $\text SU(\nu)$ basis is summarized.
We should define a regularization for the U(1) current
which preserves the local $\text U_V (1)$ gauge symmetry
$\psi_{L} \rightarrow  \exp{({\text i}\alpha)} \psi_{\!{\scriptscriptstyle 
L}},\quad \psi_{R} \rightarrow \exp{-({\text i}\alpha)}
\psi_{\!{\scriptscriptstyle R}}$.
An arbitrary local composite operator 
should be defined
in the gauge invariant point spritting regularization
\cite{IM}.
The current operators are defined in this way as well.
We shall discuss  the importance of the marginal perturbation
${\cal L}^{\ }_{\rm int}$ on the basis of
a RG calculation later.
To solve this system ${\cal L}_0$, we take a gauge fixing condition $A=0$
and a parametrization of $\overline{A}$ with a
scalar field $\phi(z,\overline{z})$
\begin{equation}
\overline{A} =\overline{\partial}\phi(z,\overline{z}), \quad
{\psi}^{\ }_{\!{\scriptscriptstyle  L}\alpha}(z)=\tilde{\psi}^{\ 
}_{\!{\scriptscriptstyle  L}\alpha}(z)\exp{(-i\phi)}.
\label{para}
\end{equation}
The gauge invariant regularization defines
the unique Jacobian for the chiral transformation
${\psi}^{\ }_{\!{\scriptscriptstyle  L}} \rightarrow
\tilde{\psi}^{\ }_{\!{\scriptscriptstyle  L}}$
which induces the kinetic term of the scalar field $\phi$ \cite{IM}.
Then, the two-dimensional abelian gauge theory with
global symmetry $\text SU(\nu)$
is expressed as a decoupled free field Lagrangians:
\begin{equation}
{\cal A}^{\ }_{\ast} = \int {d^{\,2}\! z \over 2 \pi} \left( {\cal L}_{\rm 
matter}+{\cal L}_{\rm gauge}+{\cal L}_{\rm ghost} \right)
\label{S0}
\end{equation}
with
$$
{\cal L}_{\rm matter} = 2\tilde{\psi}_{\!{\scriptscriptstyle  L 
}\alpha}^{\dagger} \overline{\partial}\tilde{\psi}^{\ 
}_{\!{\scriptscriptstyle  L}\alpha}+2\psi_{\!{\scriptscriptstyle 
 R}\alpha}^{\dagger}\partial \psi^{\ }_{\!{\scriptscriptstyle  R}\alpha},
$$
\begin{equation}
{\cal L}_{\rm gauge} = - \nu \partial \phi \overline{\partial} \phi ,
\label{mgg}
\end{equation}
$$
{\cal L}_{\rm ghost} = 2\overline{\eta}\partial \overline{\epsilon}+ 
2\eta\overline{\partial}\epsilon,
$$
where the last term is the Fadeev-Popov ghost one originating from
the measure of the gauge field parametrizedin terms of $\phi$.
Hereafter,  the tilde of the left moving fermions
will be omitted for simplicity.
The operator product expansions (OPE) between
free fields are
$$
\psi^{\dagger}_{\!{\scriptscriptstyle  L}\alpha}(z) \psi^{\ 
}_{\!{\scriptscriptstyle  L}\beta}(\omega) \sim \displaystyle 
 {\delta_{\alpha,\beta}\over z-\omega} + \cdots,
$$
$$
{\psi}^{\dagger}_{\!{\scriptscriptstyle  R}\alpha}(\overline{z}){\psi}^{\ 
}_{\!{\scriptscriptstyle  R}\beta}(\overline{\omega}) \sim \displaystyle 
 {\delta_{\alpha,\beta}\over \overline{z}-\overline{\omega}} + \cdots,
$$
\begin{equation}
e^{{\text i}\phi(z,\overline{z})}e^{-{\text 
i}\phi(\omega,\overline{\omega})} \sim |z-\omega|^{2/\nu}_{\ } + \cdots
\label{fope}
\end{equation}
which allow us to calculate OPE for energy-momentum tensors
and read off the central charges
$$
c_{\rm matter}=\nu,\quad c_{\rm gauge}=1,\quad c_{\rm ghost}=-2.
$$
As expected,  the total central charge $c_{\rm total}$ equals to `$\nu-1$' 
which agree with the Bethe ansatz's result \cite{deV}.
The negative sign in the Lagrangian
${\cal L}_{\rm gauge}$, eq(\ref{mgg}), suggests
that  the U(1) degrees of freedom
freeze in the asymptotic behaviors  of the spin system.
Since the conformal weight of the vertex operator becomes negative,
eq(\ref{fope}) shows unphysical infrared behavior, and thus
it should not appear by itself.
Actually, the U(1) current regularized
in the gauge invariant way 
\begin{equation}
j^{\ }_{\!{\scriptscriptstyle L}}(z)=
\,:\tilde{\psi}^{\dagger}_{\!{\scriptscriptstyle  L }\alpha}
(z) \tilde{\psi}^{\ }_{\!{\scriptscriptstyle  L}\alpha}(z):\
\label{AU1}
\end{equation}
(\ref{AU1})
has no Goto-Imamura-Schwinger term
in the U(1) Kac-Moody algebra.
Therefore, this U(1) Kac-Moody algebra has only a trivial
representation $j_{\!{\scriptscriptstyle L}}=0,
j_{\!{\scriptscriptstyle R}}=0$.
According to the bosonization formula in
the ${\text U}(\nu) = {\text U}(1) \times {\text SU}(\nu)$-invariant
free Dirac theory, degrees of freedom of
U(1)(charge) and ${\text SU}(\nu)$ (spin)
are separated  into a scalar boson and a $\text SU(\nu)$ WZW theory.
In our case, the U(1) degree of freedom is killed by the gauge field
which is represented by the scalar boson with
negative norm.
Eq(\ref{S0}) can be
identified with the level-one $\text SU(\nu)$ WZW model \cite{Poly}.
The Wess-Zumino primary field $G(z,\overline{z})$ is given by
$$
G_{\alpha\beta}^{\ }(z,\overline{z}) \propto 
\psi^{\dagger}_{\!{\scriptscriptstyle L}\alpha}(z)e^{ {\text 
i}\phi(z,\overline{z})}\psi^{\ }_{\!{\scriptscriptstyle 
R}\beta}(\overline{z}),\quad G(z,\overline{z})\in {\text SU({\it \nu})}
$$
and the conformal weight is $(\nu-1)/2\nu$.
To compute the asymptotic behavior
of the spin correlation function for the bulk,
it is enough to replace the spin-one operators
(\ref{SPINop}) by the continuum fields
in terms of eq(\ref{Lfermion}) and eq(\ref{para})
$$
a^{-1}_{\ }\bm{S}^{\ }_{j} \simeq \bm{J}_{\!\scriptscriptstyle L}(r) + 
\bm{J}_{\!\scriptscriptstyle R}(r) + \left[  e^{ {\text 
i}\phi(r)}\psi^{\dagger}_{\!{\scriptscriptstyle L}\alpha}(r)\bm{L}^{\ 
}_{\alpha\beta} \psi^{\ }_{\!{\scriptscriptstyle R}\beta}
(r) \exp{(2{\text i}k_F r)} + {\rm h.c} \right],
$$
where $\bm{J}^{\ }_{\!\scriptscriptstyle 
L(R)}(r)=\psi^{\dagger}_{\!\scriptscriptstyle 
L(R)\alpha}(r)\bm{L}^{}_{\alpha\beta}\psi^{\ }_{\!\scriptscriptstyle 
L(R)\beta}(r)$.
Using this, we obtain the typical correlation function of Tomonaga-Luttinger 
liquids \cite{KY}
\begin{equation}
\langle\bm{S}^{\ }_r\cdot \bm{S}^{\ }_{0} \rangle \propto  {1\ \over\ r^2} 
+{\rm constant} \times \frac{\cos{(2k_F r)}}{r^{2x}_{\ }}
\label{Scorre}
\end{equation}
with scaling dimension $x= 1-1/\nu$, in which the second term is dominant as 
$r\rightarrow \infty$ and the momentum distribution shows a power law 
singularity  near the fermi momentum $k_F$.
The appearance of the oscillating factor is reflection of
the chiral $\bm{Z}_{\nu}$ symmetry in the antiferromagnet.

\setcounter{equation}{0}
\section{The role of marginal operators}
\label{sec:marginal}
We have neglected the marginal operators  so far.
One of them is  the $\text SU(\nu)$ current interaction
in ${\cal L}^{\ }_{\rm  int}$ which gives logarithmic finite-size 
corrections.
Besides, there is another operator
$\sum_{j=1}^Lc^{\dagger}_{j,\alpha}c^{\ 
}_{j,\beta}c^{\dagger}_{j+1,\alpha}c{\ }_{j+1,\beta}(x)$
which breaks the global $\text SU(\nu)$ symmetry
except at the ULS point.
The continuum form  of the  $\text SU(\nu)$-asymmetric interaction  is given 
in terms of eq(\ref{Lfermion}) by
$$
c^{\dagger}_{j,\alpha}c^{\ }_{j,\beta}c^{\dagger}_{j+1,\alpha}c^{\ 
}_{j+1,\beta} \simeq \psi^{\dagger}_{\!{\scriptscriptstyle L}\alpha}\psi^{\ 
}_{\!{\scriptscriptstyle L}\beta}\psi^{\dagger}_{\!{\scriptscriptstyle 
R}\alpha}\psi^{\ }_{\!{\scriptscriptstyle R}\beta} + \cdots .
$$
which is also chiral $\bm{Z}^{\ }_{\nu}$-invariant.
The corresponding field theory is
expressed by the WZW model with these marginal perturbations
without global $\text SU(\nu)$ symmetry.
This $\text SU(\nu)$-breaking operator
becomes marginally relevant for the coupling constant $\gamma < 1$,
and thus a dynamical mass generation is expected.

We consider a perturbed CFT with the following action:
\begin{equation}
{\cal A} = {\cal A}^{\ }_{\text SU(\nu)^{\ }_1} + \sum_{i=1}^{2} g^{\ }_{i} 
\int \frac{d^{\,2}\! z}{2\pi} \Phi^{(i)}_{\ }(z,\overline{z}) ,
\label{Ac}
\end{equation}
where
\begin{equation}
\begin{array}{@{\,}c@{\ }c@{\ }l}
\Phi_{\ }^{(1)}(z,\overline{z}) &=& \displaystyle 
{2\over\sqrt{\nu^2-1}}{\cal J}^A_{\!\scriptscriptstyle L}(z) {\cal 
J}^A_{\!\scriptscriptstyle R}(\overline{z}), \medskip \\
\Phi_{\ }^{(2)}(z,\overline{z}) &=& \displaystyle {4 
T^A_{\alpha\beta}T^B_{\alpha\beta}\over \sqrt{\nu^2-1}}{\cal 
 J}^A_{\!\scriptscriptstyle L}(z) {\cal J}^B_{\!\scriptscriptstyle 
R}(\overline{z}).
\end{array}
\label{Phi12}
\end{equation}
There are no other relevant or marginal operators
with rotational and chiral $\bm{Z}_3$ symmetry.
The coupling constant $g_{2}$ is proportional to
$\gamma-1$  with a positive coefficient in the case of $\nu=3$ ($S=1$).
The unperturbed action ${\cal A}^{\ }_{\text SU(\nu)^{\ }_1}$
is  given by eq(\ref{S0}) and the marginal operators obey the OPE algebra
$$
\Phi^{(1)}_{\ }(z,\overline{z})\Phi^{(1)}_{\ }(0,0) \sim \displaystyle {1 
\over \ |z|^{4} } - {b \over \ |z|^2} \Phi^{(1)}_{\ }(0,0)+ \cdots,
$$
$$
\Phi^{(2)}_{\ }(z,\overline{z})\Phi^{(2)}_{\ }(0,0) \sim \displaystyle {1 
\over \ |z|^{4}}+{b \over \ |z|^2 } \Phi^{(2)}_{\ }(0,0)+ \cdots ,
$$
$$
\Phi^{(1)}_{\ }(z,\overline{z})\Phi^{(2)}_{\ }(0,0) \sim \displaystyle {1 
\over  (\nu+1) }{1\over \ |z|^{4}} - {\tilde{b} \over \  |z|^2} 
\left(\Phi^{(1)}_{\ }(0,0)-\Phi^{(2)}_{\ }(0,0)\right) + \cdots ,
$$
where
$$
b = {2\nu \over \sqrt{\nu^2 -1}},\qquad \tilde{b} = {2 \over \sqrt{\nu^2 
-1}}.
$$
This algebra gives the following one-loop $\beta$ functions
\begin{equation}
\begin{array}{@{\,}c@{\ }c@{\ }l}
\displaystyle \beta_{1}^{\ }(g_{1}^{\ } ,g_{2}^{\ }) \equiv {d g_{1}^{\ } 
\over dl} &=& \displaystyle {b \over 2} {g_{1}^{\ }}^{2} +
\tilde{b}  g_{1}^{\ }g_{2}^{\ } + O({g_{1}^{\ }}^3,{g_{2}^{\ }}^2), \medskip 
\\
\displaystyle \beta_{2}^{\ }(g_{1}^{\ },g_{2}^{\ }) \equiv {d g_{2}^{\ } 
\over dl} &=& \displaystyle - {b \over 2} {g_{2}^{\ }}^{2} -\tilde{b} 
 g_{1}^{\ }g_{2}^{\ } + O({g_{2}^{\ }}^3,{g_{1}^{\ }}^2),
\end{array}
\label{RGEQ}
\end{equation}
where $e^{l}=a$.
These coupled differential equations can be solved in
an integral form thanks to a conservation law.
An arbitrary trajectory in the coupling constant
space $(g^{\ }_1,g^{\ }_2)$
obeys the following  equation:
\begin{equation}
X^2 -Y^2  = C |Y|^{(\nu-2)/\nu}
\label{RGEsol}
\end{equation}
with $X=g_{1}^{\ }-g_{2}^{\ }$ and $Y= - g_{1}^{\ } - g_{2}^{\ }$,
where $C$ is an arbitrary real constant.
The sign of the initial value of $g_{1}^{\ }$ should be chosen
to be negative
in order to agree with the result of the Bethe ansatz.
The running coupling constant
$g_{1}^{\ }$ is renormalized to be zero in the infrared limit.
Therefore this model is in the second region
$g_{1}^{\ } < 0, g_{2}^{\ } > 0$or
the third one $g_{1}^{\ }< 0, g_{2}^{\ } < 0 $ in the coupling constant 
space.
In the $\nu=2$ case, this perturbed CFT describes
the well-known spin-$1/2$ XXZ chain and eq(\ref{RGEsol})
shows a hyperbolic trajectory where the BKT transition
occurs beyond the $\text SU(2)$ symmetric line $X \pm Y = 0 $.
There is one parameter family of fixed points
in $c^{\ }_{\rm vir}=1$ ,
that is  a fixed line
$g_1 + g_2 = 0 $ \cite{Ts}.
Note that the topology of the flow diagram in
the case of $\nu \neq 2$ differs from that in $\nu =2$.
The only fixed point is $g_{1}^{\ast}=g_{2}^{\ast}=0$
except for $\nu = 2$.

Here, we show  main results of the RG flow
which will be illustrated in the remaining part of this paper.
The RG argument classifies the coupling constant space
with $g_1 < 0$ into the following three cases:
\begin{itemize}
\item $g_2 =0$; $\text SU(\nu)$ symmetric and asymptotically non-free,
\item $g_2 > 0$; $\text SU(\nu)$ asymmetric and asymptotically non-free,
\item $g_2 < 0$; $\text SU(\nu)$ asymmetric and asymptotically free.
\end{itemize}
Since the interaction $\Phi ^{(2)}$ is marginally relevant for $g_2 < 0$,
and is marginally irrelevant for $g_2 > 0$,
the trajectory along $g_{2}^{\ } = 0$
becomes the BKT transition line.
In a finite system in the asymptotically non-free region $g_2 \geq 0$,
thermodynamic quantities acquire
some corrections due to the presence of
marginally irrelevant operators,
while in an infinite volume limit
there is no influence from them.
We indicate the difference of the finite-size corrections between
$\text SU(\nu)$- symmetric and asymmetric models.
In the third case of $g_2 < 0$,
the marginally relevant
interaction $\Phi ^{(2)}$ can generate a mass gap which might be
interpreted as the Haldane gap.

First, following Ludwig and Cardy \cite{LC,Ca}, we calculate finite-size 
corrections in the $\text SU(\nu)$ symmetric model ($g_2 =0$) with $g_1 <0 
$.
The finite-size corrections to the ground state energy
of the $\text SU(\nu)$-symmetric models in the
eq(\ref{Genergy}) are calculated as

\begin{equation}
c^{\ }_{\rm vir}=\nu-1,\quad d^{\ }_{\rm G.S}
=\frac{\nu^2_{\ } -1}{2\nu^2_{\ }},
\label{cd}
\end{equation}
where $\nu=2S+1$ for the spin-$S$.

The finite-size corrections to the low-lying
excited energies are calculated from
the most relevant primary field
$$
{\cal O}^{A}_{\ }(z,\overline{z}) = \psi^{\dagger}_{\!{\scriptscriptstyle 
L}\alpha}(z)T^{A}_{\alpha \beta} \psi^{\ }_{\!{\scriptscriptstyle 
R}\beta}(\overline{z}) e^{ {\text i}\phi(z,\overline{z})} + {\rm h.c},
$$
where $T^{A}$'s for $A =1,\cdots,
\nu^2-1$ are the $\text SU(\nu)$ basis,
$T^{0}_{\ } \equiv I / \sqrt{2\nu} $,
and they are also normalized as ${\rm Tr}[T^AT^B]=\delta_{\ }^{AB}/2$.
The primary states, $|{\cal O}^A_{\rm in}\rangle
\equiv \lim_{ z,\overline{z}\rightarrow 0}{\cal O}^A(z,\overline{z})|
0\rangle$,  become  eigenstates
of Virasoro's charge $L_0^{\ }$
$(\overline{L}_{0}^{\ })$ with an eigenvalue $x/2$.
Their OPE  are given by
$$
{\cal O}^A_{\ }(z,\overline{z}) {\cal O}^B_{\ }(0,0) \sim  \displaystyle 
{\delta^{AB}_{\ }\over \ |z|_{\ }^{2-2/\nu}} + \cdots ,
$$
$$
{\cal O}^A_{\ }(z,\overline{z})\Phi^{(1)}_{\ }(0,0) \sim \displaystyle 
-{b^{\ }_{A}\over\  |z|_{\ }^{2}} {\cal O}^A_{\ }(0,0)+ \cdots
$$
with the OPE coefficients
$$
b_{A}^{\ } = {1 \over \nu \sqrt{\nu^2-1} }
\times \left\{ \begin{array}{@{\,}c@{\ }c@{\ }l}
{\nu^2-1}  &\ &  {\rm for}\  A = 0,\medskip \\ -  1
&\  & {\rm for}\ A = 1,\cdots,\nu^2-1 .\end{array} \right.
$$

We obtain the universal quantities in eq(\ref{Exenergy})
\begin{equation}
x_{A}^{\ } = 1 - 1/\nu, \quad
d_{A}^{\ } ={2b_{A}^{\ }\over b} =\left\{ \begin{array}{@{\,}c@{\ }c@{\ }l} 

1 -  1/\nu^{2}_{\ } &\ & {\rm for}\
A = 0 \medskip \\ - 1/\nu^{2}_{\ } &\ & {\rm for}\
A= 1, \cdots  , \nu^2-1. \end{array} \right.
\label{xd}
\end{equation}
These $\nu^2$ states are classified  by the total spin.
As shown in Appendix A, the state $A=0$ describes the singlet
excitation and other
$\nu^2-1$ primary states are higher spin states with spin up to
$(\nu -1)/2$.
In the finite-size corrections up to the logarithmic size dependence,
the singlet excitation is not favored
compared to those with higher spin.

The finite-size effect for the spin correlation function
(\ref{Scorre}) is obtained immediately
from the information on the excited energy \cite{HHM,ABZ,GS}.
\begin{equation}
\langle\bm{S}^{\ }_{r}\cdot \bm{S}^{\ }_{0} \rangle
\approx \cos{(2k_{F}^{\ }r)}{\cal G}_{A}^{\ }(g_{1}^{\ }(r),r),\quad
{\cal G}_{A}^{\ }(g_{1}^{\ }(r),r) =
\displaystyle {\  {(\ln{r})}^{\sigma_{\!A}^{\ }}
\over r^{2x_{\!A}^{\ }}_{\ }},
\label{green}
\end{equation}
where
$$
\sigma_{\!A}^{\ }=-2d_{A}^{\ }= {2 \over \ \nu^{2}_{\ }}
$$
except for A=0.
Our results for $\nu=2$ listed in Table \ref{Table2}
agrees with the Bethe ansatz's ones in ref\cite{AGSZ,WE}.
The leading finite-size corrections $c_{vir}$ and $x_A$
in the SU($\nu$)symmetric model
agree with Bethe ansatz \cite{deV, PT}, as well.

Now we consider the second case $g^{\ }_{2} >0$, where
there is a marginally irrelevant $\text SU(\nu)$-asymmetric interaction.
The situation  is crucial whether $\nu=2$ or not.
Even though the action describing the ultraviolet
theory has no $\text SU(\nu)$ symmetry due to
the $\text SU(\nu)$-breaking interaction,
the $\text SU(\nu)$-breaking interaction
have no effect on the leading terms of the finite-size correction
except in $\nu=2$ case.
The RG indicates that the $\text SU(\nu)$ symmetry appears dynamically
for macroscopic scale even though the $g_{2}^{\ }$-term
in eq(\ref{Phi12}) is switched into the fixed point action.
The difference between the $\text SU(\nu)$-symmetric and
asymmetric model appears in the logarithmic correction term.

To calculate the logarithmic correction,
we note that
the RG flow eq(\ref{RGEQ})
with an initial condition  $g_{1}^{\ }<0$ and  $g_{2}^{\ }>0$
is absorbed into the fixed point along the line
\begin{equation}
g_{1} = - g_{2}.
\label{g1g2}
\end{equation}
The macroscopic property of
the system is determined by the scale $l \gg 1$,
and we can estimate the deviation from the line as
 $|g_1 (l) + g_2 (l)| \sim O(l^{-\frac{2 \nu}{\nu - 2}})$
with the help of the integral curve eq(\ref{RGEsol})
for arbitrary solution with an initial condition in the second region.
Therefore we can calculate the logarithmic correction
by assuming that the marginally
irrelevant flow for $g_{2}^{\ }>0$ is described by the action
$$
{\cal A} = {\cal A}^{\ }_{\text SU(\nu)^{\ }_1} + g^{\ }_{1} \int 
\frac{d^{\,2}\! z}{2\pi} \Psi(z,\overline{z})
$$
with ${\Psi}(z,\overline{z}) = \sqrt{\nu+1\over2\nu}\left(\Phi_{\ 
}^{(1)}(z,\overline{z})- \Phi_{\ }^{(2)}(z,\overline{z})\right)$ which is 
normalized by
$$
\Psi(z,\overline{z})\Psi(0,0) \sim {1\over \ |z|_{\ }^{4}}-{B \over \ 
|z|^2_{\ }}\Psi(0,0),
$$
where the OPE coefficient $B$ is

$$
B=\sqrt{\nu + 1 \over 2\nu} ( b-2\tilde{b} ).
$$
This assumption might hold, since the current of
the RG  would spend the fair time near the fixed point with dilatation.
As in the discussions of the symmetric model,
we can  evaluate the coefficients of
the finite-size energy correction from the one-loop
renormalization which obeys
$ d g_{1}^{\ } / dl =  (B/2 ){g_{1}^{\ }}^{2} $.
For the ground state energy, we obtain
$$
c^{\ }_{\rm vir}=\nu-1, \quad
d_{\rm G.S}^{\ }= \frac{\nu(\nu-1)}{\ (\nu-2)^2_{\ }},
$$
in which  the logarithmic coefficient is different from  eq(\ref{cd}).
The three-point function in the expression of the excited energy
is given by using the OPE
$$
{\cal O}^A_{\ }(z,\overline{z})\tilde{\Phi}(0,0) \sim  -{B^{\ }_{A}\over\ 
 |z|_{\ }^{2}} {\cal O}^A_{\ }(0,0)+ \cdots.
$$
Here the coefficient $B_{A}^{\ }$ takes three different values  according to 
the symmetric properties of the matrices $\{ T^{A}_{\ } \}$
under the matrix transposition.
These are given by
$$
B_{A}^{\ }={1 \over \sqrt{2\nu(\nu-1)} }\times\left\{
\begin{array}{@{\,}c@{\ }c@{\ }l}
\nu-1\ &{\rm for}&\ A=0\\ 1\  &{\rm for}&\ A(\neq0)\ {\rm with} {\ }^{\rm 
t}(T^{A}_{\ }) = - \ T^{A}_{\ } \\ -1\  &{\rm for}&\  A(\neq0)\ {\rm with} 
{\ }^{\rm t}(T^{A}_{\ }) =T^{A}_{\ }.  \end{array}\right.
$$
As a result, we have the universal coefficients in the anomalous dimension 
(\ref{Exenergy})
\begin{equation}
x_{A}^{\ } = 1 - 1/\nu, \quad
d_{A}^{\ } = {1 \over \nu-2} \times \left\{ \begin{array}{@{\,}c@{\ }c@{\ 
}l}
 \nu - 1  &{\rm for}&\  A = 0  \\
 1  &{\rm for}&\ A(\neq0)\ {\rm with} {\ }^{\rm t}(T^{A}_{\ }) = - \ 
T^{A}_{\ } \\ - 1   &{\rm for}&\ A(\neq0)\ {\rm with} {\ }^{\rm t}(T^{A}_{\ 
}) =  \ T^{A}_{\ }.
\end{array} \right.
\end{equation}
The OPE coefficents $B$ and $B_{A}^{\ }$ give the exponents 
$\sigma_{\!A\neq0}^{\ }$ characterizing logarithmic distance dependence in 
eq(\ref{green})
$$
\sigma_{\!A}^{\ } = {2 \over (\nu-2)}.
$$
The primary states with $A\neq0$ in the symmetric model are degenerate even 
if we consider the logarithmic correction,
those in the asymmetric one splits to two levels.
As shown in Appendix A, the difference of the OPE coefficients because of 
 the symmetric and antisymmetric
properties of $\text SU(\nu)$ Lie algebra  basis is classified by total 
spin.
In particular, for the $\text SU(3)$ problems,
the primary with the identity matrix (A=0) is spin-singlet, three primaries 
with the antisymmetric matrices are spin-triplet and the remainder with 
symmetric ones are spin-quintuplet.
The universal coefficients characterizing  the $\text SU(3)$-symmetric
and asymmetric model are shown in Table \ref{Table3}.

Let us now consider the third case
$g_{2}^{\ }<0$, which corresponds to $\theta < \theta_c \equiv
\pi/4$ in the $S=1$ model.
The theory is asymptotically free, 
then we expect the mass generation which can be identified with
the Haldane gap in the  $S=1$ case.
One can estimate the mass gap by solving
the renormalization group eq(\ref{RGEQ}).
The conservation law eq(\ref{RGEsol}) enables us to reduce
the simultaneouse equation for the two unknown
functions $g_1$ and $g_2$ to that for the one unknown $Y = -g_1 -g_2$
\begin{equation}
\frac{dY}{dl} = \pm \nu \sqrt{\nu^2 -1} Y^2 \sqrt{1 + C Y^{-(\nu+2)/\nu}},
\label{1PRGE}
\end{equation}
where the sign of the right hand side is identical to
that of $X = g_1 -g_2$.
Let us set the initial condition of the running coupling
constants near the transition point
$$
g_1 (0) = - a_1 \quad g_2 (0) \simeq a_2( \theta -\theta_c),
$$
where $a_1$ and $a_2$  are positive constants.
This condition sets the integral constant as $C \simeq a_3 (\theta - 
\theta_c)$
in eq(\ref{RGEsol}) with a positive constant $a_3$.
The renormarization group equation eq(\ref{1PRGE}) is immediately
integrated under this condition
\begin{equation}
\left( -\int_{Y(0)} ^{|C|^{\sigma}} +
\int_{|C|^{\sigma}} ^{Y(l)} \right)
\frac{d Y}{Y^2 \sqrt{1+ C Y^{-1/\sigma}}}
= \nu \sqrt{\nu^2 -1} \ln l,
\label{RGEint}
\end{equation}
where $\sigma \equiv \frac{\nu}{\nu + 2}$.
This gives us the order of the scale $m^{-1}$ which makes the running
coupling constant diverge $g_2 (\ln m^{-1}) = \infty $
This scale $m$ is the energy gap
\begin{equation}
m = \exp{\left( - A |C|^{-\sigma} \right)}
\simeq \exp{\left(-c (\theta_c - \theta)^{-\sigma} \right)},
\label{gap}
\end{equation}
where
$$
A = \frac{2}{\nu \sqrt{\nu^2-1}} \int_1 ^\infty
\frac{dy}{y^2 \sqrt{1-y^{-1/\sigma}}},
$$
and $c = a_3 A $ is positive.
Therefore we conclude that
the phase transition is infinite order.
This result agrees with the recent numerical studies of $S=1$ model by
F\'ath and S\'olyom \cite{FS}. To see this, one should check
their obtained energy gap directly rather than one
parameter beta function estimated from it, since we have
two parameter beta function (\ref{RGEQ}).
Their numerical data of the energy gap fit the function eq(\ref{gap})
with the universal constant $\sigma = 0.8 \pm 0.2$.
This is consistent with
our result $\sigma = \frac{\nu}{\nu+2} = 0.6 $ at $\nu=3$.

\section{Discussion and Open problems}
We have investigated the isotropic spin-1 model to clarify
the phase diagram around the Uimin-Lai-Sutherland (ULS) point.
The low energy  theory of the ULS model is described
by a strong coupling abelian gauge theory
which can be regarded as the critical level-one ${\text SU}(3)$ WZW model. 

We have shown a mechanism of the dynamical mass generation in the $S=1$
Haldane phase in the presence of the ${\text SU}(3)$
breaking interaction with dimension 2.
We have shown that the dimension 2 operator makes
the massless phase $\theta \leq \pi / 4$ and
the massive phase $\theta < \pi / 4 $ around the ULS point $\theta = \pi / 4 
$
in the model eq(\ref{Hamil}).
This nature can be understood by the level $k=1$ WZW theory, which
has neither relevant operator with the chiral $\bm{Z}_3$ invariance
nor tensored operator of the WZ matrices but
merely marginal operators.
Therefore, the Haldane phase has the exponential mass gap
as a result of the BKT transition. The region $\pi /2 \leq \theta \leq \pi 
/4$
is concluded to be massless from this analysis and the
numerical study \cite{FS}.
Here, we indicate the difference of the phase transitions
at the ULS point and at another integrable point
$\theta= -\pi/4$ of the Takhatajan-Babujian (TB) model.
In alternative field theoretical approach for understanding the Haldane 
massive phase, Affleck and Haldane investigated the relevant
deformation of the $S=1$ T-B model \cite{AH}.
The universality class of this T-B model is the level-two $\text SU(2)$
WZW model, where the one-site translation  corresponds to the chiral
$\bm{Z}^{\ }_{2}$ transformation.
In the level-$k$ theory with $k > 1$,
one can make the chiral $\bm{Z}_2^{\ }$ invariant relevant operator in terms 
of tensoring of the SU(2) WZ matrices $G(z,\overline{z})$, for example 
$({\rm Tr}[G])^2$.
Therefore the transition from that massless point to the Haldane phase
becomes second order, and the mass gap opens obeying the power law.
In this case, the T-B point $\theta = -\pi / 4 $ is isolated
as a massless point in the massive region, namely the
Haldane phase $\theta > -\pi / 4 $ and the dimer phase
$\theta < -\pi / 4 $.

The renormalization group flow given by eq(\ref{RGEQ})
has a unique fixed point in $\nu > 2$
case, while that in $\nu = 2 $ case has a fixed line.
Contrary to the $\nu = 2$ case, the logarithmic corrections appears
in the massless phase for $\nu > 2$ even if
there is the $\text SU(\nu)$ symmetry breaking interaction.
We have calculated coefficients of logarithmic corrections
to the energies of the ground state and some excited states
both in $\text SU(\nu)$ symmetric and asymmetric models.
We find the different coefficients in these two cases.from their numerical 
data of the energy gap as in the form eq(\ref{gap})
This nature of the model with $\nu > 2$ suggests
Cardy's argument that a natural irreducible CFT with one parameter
should have the central charge $c_{\rm vir}^{\ }=1$ \cite{Ca2}.

Nonetheless, no one has ever succeeded in classifying
CFT with $c_{\rm vir}^{\ }>1$, and therefore
to search CFT with fixed line (or surface) might be worth attempting.
Since we need to spread the coupling constant space at least,
the simplest candidate
is a model with
anisotropic parameters or q-deformation of
the Lie algebra $\text SU(\nu)$.  This program is now in progress.

Here we present  some conjectures deduced from the  CFT kinematics.
We note that $\bm{J}^{\ }_{\!\scriptscriptstyle L}(z)$ and
$\bm{J}^{\ }_{\!\scriptscriptstyle R}(\overline{z})$ which are in a 
subalgebra
of ${\text SU}(3)_1$ Kac-Moody algebra except the normalization
satisfy the level-four $\text SU(2)$ Kac-Moody algebra.
The representation of
${\text SU}(3)_1^{\ }$ is involved in
that of ${\text SU}(2)_4^{\ }$.
The  central charge of both theories are $c_{\rm vir}^{\ }=2$ and conformal 

weight of the primary field with spin-$j$ is
$\triangle^{(j)}_{\ }=j(j+1)/6$ with $0 \leq j \leq 2$ \cite{ZF}.
If we neglect primaries with half-odd-integer spin
in the ${\text SU}(2)_4$ WZW model, we obtain those in the ${\text SU}(3)_1$ 
WZW model.
The ${\text SU}(2)_{4}^{\ }$ WZW model can be regarded as a critical theory 
of
the spin-$2$ TB model, and therefore we can expect the following prediction:

\begin{description}
\item{Conjecture 1} {\it There is a cross-over flow
from the spin-$2$  Takhtajan-Babujian model to the spin-$1$
Uimin-Lai-Sutherland model.}
\end{description}

As recognized in the studies of the $\text SU(2)$ spin chains, coefficients 

$d_j$ in the logarithmic correction to
the excited states with total spin-$j$ satisfy the following
sum rule \cite{ZS}: $3 d_t+ 1 d_s=0$,
where $d^{\ }_{s(t)}$ is the universal coefficient for the singlet (triplet) 

excitation(s) and the prefactor is the dimension of the spin 
representation.
We have seen that such a similar rule exists in the spin-1 models discussed 

above, as well.
That is $5d^{\ }_q+3d^{\ }_t+d^{\ }_s=0$  \cite{No}.
Therefore, we are led to the following conjecture:

\begin{description}
\item{Conjecture 2} {\it There exists a sum rule among  the coefficients 
 $\{d_j^{\ }\}$ of the leading
logarithmic correction term in the excited energy with total spin-$j$; 
i.e}.
$$
\sum_{j=0}^{2S}(2j+1)d_{j}^{\ }= 0.
$$
\end{description}

\acknowledgments
We thank S. Hikami, A. Kitazawa, H. Mukaida, K. Nomura and K. Okamoto for
helpful discussions.
We are grateful to  T. Fujita and S. Misawa for carefully
reading the  manuscript.
The research of M. -H. K is supported in part by
a JSPS Research Fellow for Young Scientists.

\newpage

\appendix
\section{}
The fundamental representation of the $\text SU(\nu)$ Lie algebra 
$[T^A,T^B]=f^{AB}_{\quad C} T^C$ is summarized  as follows.
The $\text SU(\nu)$ exchange operator is decomposed in terms of $\text 
SU(\nu)$ basis as
$$
{\cal P} = {1 \over \nu} I \times I + 2\sum_{A=1}^{\nu_{\ }^2-1}T^A \times 
T^A. $$
These basis are  normalized as ${\rm Tr}[T^AT^B]={1\over 2}\delta^{AB}$ or 

$\sum_{A=1}^{\nu^2-1}T^AT^A=(\nu^2-1)/2\nu$.
The  structure constant $f^{ABC}_{\ }$ has the quadratic Casimir of the 
adjoint representation: 
 $\sum_{A,B=1}^{\nu^2-1}f^{ABC}f_{ABD}=-{\nu}\delta^{C}_{\ \ D}$.
Another expression of the exchange operator is available when the spin 
chains are studied.
On a space $\bm{C}^{2S+1} \times \bm{C}^{2S+1}$, it is  given by
$$
{\cal P} = {(-1)}_{\ }^{2S}\sum_{j=0}^{2S}(-1)^j_{\ }{\cal P}_{\ }^{(j)},
$$
where ${\cal P}^{(j)}$ is the projector onto a space of spin-$i$  conforming 
to  an identity $I \times I = {\cal P}^{(0)}+\cdots + {\cal P}^{(2S)}$.
The projector ${\cal P}^{(j)}$ on a spin-$j$ space is represented using the 
spin operators with the magnitude $S$ as follows:
$$
{\cal P}^{(j)} = \prod_{\stackrel{\scriptstyle k=0}{ (\neq j)}}^{2S}\left[ 
{\ {\text X}-x_k \  \over x^{\ }_j - x^{\ }_k } \right], \quad {\text X} = 
\sum_{a=1}^{3}S^a \times S^a,
$$
where $x_k=[k(k+1)-2S(S+1)]/2$.
The  expressions of the exchange operator in terms of the spin operator
are shown in Table \ref{Table1}.

In particular, the representation of $\text SU(3)$ is realized  by Gell-Mann 
matrices
$$
\begin{array}{@{\,}c@{\ }c@{\ }l}
\lambda_1^{\ }=\left( \begin{array}{ccc}
0&0&1 \\ 1&0&0 \\ 0&0&0 \end{array}
\right),\quad
\lambda_2^{\ }=\left( \begin{array}{ccc}
0&-i&0 \\ i&0&0 \\ 0&0&0 \end{array}
\right),\quad
\lambda_3^{\ }=\left( \begin{array}{ccc}
1&0&0 \\ 0&-1&0 \\ 0&0&0 \end{array}
\right), \medskip \\
\lambda_4^{\ }=\left( \begin{array}{ccc}
0&0&1 \\ 0&0&0 \\ 1&0&0 \end{array}
\right),\quad
\lambda_5^{\ }=\left( \begin{array}{ccc}
0&0&-i \\ 0&0&0 \\ i&0&0 \end{array}
\right),\quad
\lambda_6^{\ }=\left( \begin{array}{ccc}
0&0&0 \\ 0&0&1 \\ 0&1&0 \end{array}
\right), \medskip \\
\lambda_7^{\ }=\left( \begin{array}{ccc}
0&0&0 \\ 0&0&-i \\ 0&i&0 \end{array}
\right),\quad
\lambda_8^{\ }=\left( \begin{array}{ccc}
1/\sqrt{3}&0&0 \\ 0&1/\sqrt{3}&0 \\ 0&0&-2/\sqrt{3} \end{array}
\right), \end{array}$$
where $T^{A}_{\ }= \lambda_{A}^{\ }/2$.
Here $\lambda^{\ }_{A=2,5,7}$ are antisymmetric matrices and the remaiders 
of them are symmetric.

The primary states $\{|{\cal O}^{A}_{\rm in}\rangle \}$ can be classified by 
 total spin-$j$.
The total spin operator is given by
$$
\bm{S}^{\ }_{\rm tot} = \int_0^L \!\!dx \ \bm{S}(x)
= \bm{J}_{\!\scriptscriptstyle L,0}^{\ }
+ \bm{J}_{\!\scriptscriptstyle R,0}^{\ },
$$
where $\text SU(2)$ charge operators  are  $\bm{J}_{\!\scriptscriptstyle 
L,0}^{\ }=\oint\!{dz\over2\pi{\text i}}\bm{J}^{\ }_{\!\scriptscriptstyle 
L}(z)$ and $\bm{J}_{\!\scriptscriptstyle R,0}^{\ }=\oint\!{d\overline{z} 
\over 2\pi{\text i}}\bm{J}^{\ }_{\!\scriptscriptstyle R}(\overline{z})$.
The magnitude of total spin of the  primary states takes values $0,1$ or $2$ 
from a synthesis of two fermions with spin 1.
Acting $\bm{S}_{\rm tot}^{\ }$ to the primary fields,
we obtain  the OPE
$$
\begin{array}{@{\,}c@{\ }c@{\ }l}
\bm{S}^{\ }_{\rm tot}{\cal O}^{A=0}_{\ }(z,\overline{z}) &=& 0, \medskip 
\\
\bm{S}^{\ }_{\rm tot}{\cal O}^{A\neq0}_{\ }(z,\overline{z}) &=& 4 {\cal 
O}^{A}_{\ }(z,\overline{z})+2{T}^{A}_{\beta\alpha}( 
\psi^{\dagger}_{\!{\scriptscriptstyle L}\alpha}(z) \psi^{\ 
}_{\!{\scriptscriptstyle R}\beta}(\overline{z}) e^{ {\text 
i}\phi(z,\overline{z})} + {\rm h.c}).
\end{array}
$$
Here we have used the properties of the $\text SU(3)$ basis.
Using the symmetric and asymmetric properties of the Gell-Mann matrices, we 
obtain
$$
(\bm{S}^{\ }_{\rm tot})^2_{\ }|{\cal O}^{A}_{\rm in}\rangle
= j(j+1) | {\cal O}^{A}_{\rm in}  \rangle,
$$
where $j=0,1,2$.
The primary with identity matrix $(A=0)$
is the singlet state ($j=0$).
Three antisymmetric ones (A=2,5,7) in the Gell-Mann matrices give 
spin-triplet states ($j=1$).
The remainders which are symmetric matrices,
become quintuplet states ($j=2$).

\begin{table}[p]
\caption{Expressions of the exchange operator for $S\leq 2$}
\begin{center}
\begin{tabular}{cl}
\qquad $S$ &$\quad {\cal P}$  \\ \hline
\qquad ${1\over2}$ & $2{\text X}+\frac{1}{2}$  \\
\qquad$1$ & ${\text X}^2 + {\text X} -1$  \\
\qquad ${3\over2}$ & $\frac{2}{9}{\text X}^3 + \frac{11}{18}{\text X}^2 - 
\frac{9}{8}{\text X} - \frac{67}{32}$  \\
\qquad $2$ & $\frac{1}{36}{\text X}^4+\frac{1}{6}{\text 
X}^3-\frac{7}{12}{\text X}^2-\frac{5}{2}{\text X}-1$  \\
\end{tabular}
\end{center}
\label{Table1}
\end{table}
\begin{table}[p]
\caption{Finite-size corrections for the spin-$1/2$ Heisenberg chain}
\begin{center}
\begin{tabular}{cccccccc}
\ &\ $c^{\ }_{\rm vir}$\ &\ $x^{\ }_{t}$\ &\ $x^{\ }_{s}$\ &\ $d^{\ }_{\rm 
G.S}$\ &\ $d^{\ }_{t}$\ &\ $d^{\ }_{s}$&\ $\sigma_{\!t}^{\ }$  \\ \hline
 ${\text SU}(2)_{1}$ WZW\ 
&1&${1/2}$&${1/2}$&${3/8}$&$-{1/4}$&${3/4}$&${1/2}$ \\
BA &1&${1/2}$&${1/2}$&$0.3433$&$-{1/4}$&${3/4}$& --- \\
\end{tabular}
\end{center}
\label{Table2}
\end{table}
\begin{table}[p]
\caption{Finite-size corrections for the spin-1 chains }
\begin{center}
\begin{tabular}{cccccccccc}
\ & \ $c^{\ }_{\rm vir}$\ & \ $x_{q}^{\ }$\ & \ $x_{t}^{\ }$\ & \ $x_{s}^{\ 
}$\ &
\ $d_{\rm G.S}^{\ }$\ & \ $d_{q}^{\ }$\  & \ $d_{t}^{\ }$\ & \ $d_{s}^{\ 
}$&
\ $\sigma_{\!q}^{\ }$\   \\ \hline
${\text SU}(3)_{1}$ WZW $(\gamma=1)$\ 
&2&${2/3}$&${2/3}$&${2/3}$&${4/9}$&$-{1/9}$&$-{1/9}$&${8/9}$&${2/9}$  \\
${\text SU}(3)_1$ WZW 
$(\gamma\neq1)$&2&${2/3}$&${2/3}$&${2/3}$&$6$&$-1$&$1$&$2$&${2}$  \\
\ \ \ \ \ \ BA&2&${2/3}$&${2/3}$&${2/3}$&---&---&---&--- &--- \\
\end{tabular}
\end{center}
\label{Table3}
\end{table}

\begin{figure}
\vspace{15cm}
\caption{The renormalization group trajectory for the $\nu>2$ model,
where $\nu=2S+1$.  The coupling $g_{2}^{\ }$ is defined in the vicinity of 
the ULS model. The BKT line $g_{2}^{\ }=0$, $g_{1}^{\ }\leq0$ corresponds to 
the pure ULS model. }
\end{figure}


\begin{references}
\bibitem{SJG} U. Schollw{\"o}ck, Th. Joli{\oe}ur and T. Garel, {\it Physical 
meaning of the Affleck-Kennedy-Lieb-Tasaki S=1 quantum spin chain} , 
cond-mat/9505086.
\bibitem{AKLT} I. Affleck, T. Kennedy, E. Lieb and H. Tasaki, Phys. Rev. 
Lett. {\bf 59},  799 (1987); Commun. Math. Phys. {\bf 115},  477 (1988); S, 
Knabe, J. Stat. Phys. {\bf 52},  627 (1988).
\bibitem{KeTa} H. Tasaki, Phys. Rev. Lett. {\bf 66} (1991) 798; T. Kennedy 
and H. Tasaki, Phys. Rev. {\bf B45}, 304 (1992).
\bibitem{ULS} G. V. Uimin, JETP Lett. {\bf 12},  225 (1970); C. K. Lai, J. 
Math. Phys. {\bf 15},  1675 (1974); B. Sutherland, Phys. Rev. {\bf B12},
3795 (1975).
\bibitem{KR} P. P. Kulish and N. Yu. Reshetikhin, Sov. Phys. JETP. {\bf 53}, 
 108 (1981).
\bibitem{TB} L. A. Takhtajan, Phys. Lett. {\bf 87A},  479 (1982); H. M. 
Babujian, Phys. Lett. {\bf 90A},  479 (1982).
\bibitem{BB} M. N. Barber and M. T. Batchelor, Phys. Rev. {\bf B40}, 4621 
(1989); M. T. Batchelor and M. N. Barber, J. Phys. {\bf A23}, L15 (1990). 

\bibitem{NB}M. P. Nightingale and H. W. Bl\"{o}te, Phys. Rev. {\bf B33}, 659 
 (1986).
\bibitem{Nomu} K. Nomura, Phys. Rev {\bf B40}, 2421 (1989).
\bibitem{Jo} H. Johannesson, Phys. Lett. {\bf A116}, 133 (1986); Nucl. Phys. 
{\bf B270}, 235 (1986).
\bibitem{PT} S. V. Pokrovski\u\i \ and A. M. Tsvelik, Sov. Phys. JETP. {\bf 
66}, 1275 (1987).
\bibitem{LS} K. Lee and P. Schlottmann, Phys. Rev. {\bf B36}, 466 (1987).
\bibitem{Sch} P. Schlottmann, Phys. Rev. {\bf B45}, 5293 (1992).
\bibitem{MNTT} L. Mezincescu, R. I. Nepomechie, P. K. Townsend and A. M. 
Tsvelik, Nucl. Phys. {\bf B406}, 681 (1993).
\bibitem{Af} I. Affleck, Nucl. Phys. {\bf B265}, 409 (1986); Nucl. Phys. 
{\bf B305}, 582 (1988).
\bibitem{BY} M. T. Batchelor and C. M. Yung, {\it Integrable SU(2)-invariant 
spin chains and the Haldane conjecture}, cond-mat/9406072; T. Kennedy, J. 
Phys. {\bf A25}, 2809 (1992).
\bibitem{deV} H. J. de Vega, J. Phys. {\bf A20}, 6023 (1987); J. Phys. {\bf 
A21}, L1089 (1988).
\bibitem{FS} G. F\'ath and J. S\'olyom, Phys. Rev. {\bf B44}, 11836 (1991); 
Phys. Rev. {\bf B47}, 872 (1993).
\bibitem{AGSZ} I. Affleck, D. Gepner, H. J. Schulz and T. Ziman, J. Phys. 
{\bf A19}, 511 (1989).
\bibitem{EAT} S. Eggert, I. Affleck and M. Takahashi, Phys. Rev. Lett. {\bf 
73}, 332 (1994).
\bibitem{HHM} K. A. Hallberg, P. Horsch and G. Mart\'inez, Phy. Rev. {\bf 
52}, R719 (1995).
\bibitem{Eg} S. Eggert, {\it Numerical evidence for multiplicative 
logarithmic corrections from marginal operators}, cond-mat/9602026.
\bibitem{ABZ} A. B. Zamolodchikov, JETP Lett. {\bf 43}, 730 (1986); Sov. J. 
Nucl. Phys. {\bf 46}, 1090 (1987).
\bibitem{LC} A. W. W. Ludwig and J. L. Cardy, Nucl. Phys. {\bf B285}, 687 
(1987).
\bibitem{Ca} J. L. Cardy, J. Phys. {\bf A19}, L1093 (1986); Erratum {\bf 
A20}, 5039 (1987).
(1985).
\bibitem{GS} T. Giamarchi and H. J. Schulz, Phys. Rev. {\bf B39}, 4620 
(1989).
\bibitem{IM} C. Itoi and H. Mukaida, J. Phys. {\bf A27}, 4695 (1994).
\bibitem{Poly} A. M. Polyakov in {\it 1988 Les Houches Lecture Note}, E. 
Brezin and J. Zinn-Justin Eds., North-Holland, Amsterdam (1990).
\bibitem{KY} N. Kawakami and S.-K. Yang, J. Phys. Condens. Matter. {\bf  3}, 
5983 (1991); Phys. Lett. {\bf A148}, 359 (1990).
\bibitem{WE} F. Woynarovich and H. -P. Eckle, J. Phys. {\bf A20}, L97 
(1987)
; C. J. Hamer, M. N. Barber and M. T. Batchelor, J. Stat. Phys. {\bf 52}, 
679 (1988).
\bibitem{ZF} A. B. Zamolodchikov and V. A. Fateev, Sov. Phys. JETP. {\bf 
43}, 657 (1986).
\bibitem{ZS} T. Ziman and H. J. Schulz, Phys. Rev. Lett. {\bf 59},140 
(1987).
\bibitem{No} A. Kitazawa, K. Nomura and K. Okamoto (private communication)
\bibitem{Ts} A. M. Tsvelik, {\it Quantum Field Theory in Condensed Matter 
Physics} (Cambridge University Press 1995).
\bibitem{AH} I. Affleck and F. D. Haldane. Phys. Rev. {\bf B36}, 5291 
(1987).
\bibitem{Ca2} J. L. Cardy, J. Phys. {\bf A20}, L891 (1987)
\end{references}
\end{document}